\documentclass[twocolumn,showpacs,preprintnumbers,amsmath,amssymb,prb,superscriptaddress]{revtex4}
\usepackage{graphicx}
\usepackage[all]{xypic}
\usepackage[english, russian]{babel}

\newcommand{\Z}{\mathbb Z}
\newcommand{\R}{\mathbb R}
\newcommand{\CC}{\mathbb C}
\newcommand{\F}{\mathcal F}
\DeclareMathOperator{\id}{id}
\DeclareMathOperator{\Ker}{Ker}
\DeclareMathOperator{\Coker}{Coker}
\DeclareMathOperator{\Image}{Im}
\renewcommand{\d}{\mathrm d}
\DeclareMathOperator{\ind}{index}

\begin{document}

\title{Zero-energy states in corrugated bilayer graphene}
\author{M. I. Katsnelson}
\email{M.Katsnelson@science.ru.nl} \affiliation{Institute for
Molecules and Materials, Radboud University of Nijmegen, 6525ED
Nijmegen, The Netherlands}

\author{M. F. Prokhorova}
\email{pmf@imm.uran.ru} \affiliation{Institute of Mathematics and Mechanics,
Kovalevskaya str. 16, 620219 Ekaterinburg, Russia}


\begin{abstract}
Anomalous quantum Hall effects in single-layer and bilayer
graphene are related with nontrivial topological properties of
electron states (Berry phases $\pi$ and 2$\pi$, respectively). It
was known that the Atiyah-Singer index theorem guarantees, for the
case of the single-layer, existence of zero-energy states for the
case of inhomogeneous magnetic fields assuming that the total flux
is non-zero. This leads, in particular, to appearance of midgap
states in corrugated graphene and topologically protects
zero-energy Landau level in corrugated single-layer graphene. Here
we apply this theorem to the case of bilayer graphene and prove
the existence of zero-energy modes for this case.
\end{abstract}

\maketitle

\section{Introduction}

Graphene, that is a two-dimensional allotrope of carbon formed by
single carbon atom sheet is a subject of hot interest now (for
review, see Refs. \onlinecite{r1,r2,r3,r4}). One of the most
interesting aspects of the graphene physics from theoretical point
of view is a deep and fruitful relation with the quantum
electrodynamics and quantum field theory
\cite{semenoff,haldane,gonzales,zitter,klein,falko,r5,collapse}.
As was proven experimentally in Refs. \onlinecite{kostya,kim}
charge carriers in the single-layer graphene are massless Dirac
fermions characterized by ``chirality'', or ``Berry phase'' $\pi$.
As a consequence, graphene demonstrates anomalous quantum Hall
effect due to existence of zero-energy Landau level. The latter
can be considered \cite{r2,r5,kostya} as a simple consequence of
the famous Atiyah-Singer index theorem \cite{AS1} which plays an
important role in the modern quantum field theory and theory of
superstrings \cite{index1,index2}.

Charge carriers in bilayer graphene, formed by {\it two} graphite
atomic sheets \cite{natphys}, can be, in a good approximation,
considered as chiral fermions with the Berry phase $2\pi$ which
leads to another type of anomalous quantum Hall effect
\cite{natphys,falko1} and to a very unusual character of electron
transmission through potential barriers \cite{klein}. Exact
solution of the Schr\"{o}dinger equation for the bilayer graphene
in homogeneous magnetic field \cite{falko1} demonstrates existence
of the zero-energy Landau level with twice larger degeneracy than
for the case of single layer. However, topological origin of this
feature was not clarified yet. This is the aim of the present
work. We will prove that the existence of the zero-energy states
in bilayer graphene is also a consequence of the Atiyah-Singer
index theorem and thus is topologically protected.

This is an important question since it is known
\cite{morozov,meyer1,meyer2,ishigami,stolyarova,fasolino} that
graphene is always corrugated and covered by ripples which can be
either intrinsic \cite{meyer1,meyer2,fasolino,nelson} or induced
by a roughness of substrate \cite{ishigami,stolyarova}. In
general, non-flatness of graphene leads to appearance of the
pseudomagnetic {\it inhomogeneous} gauge field
\cite{morozov,morpurgo} acting on the charge carriers. Based on
the topological arguments one can demonstrate that this
pseudomagnetic field should also result in appearance of
zero-energy states (pseudo-Landau levels) as was recently
confirmed by model \cite{guinea} and first-principles \cite{tim}
electronic structure calculations. This can provide a mechanism of
formation of charge inhomogeneity in corrugated graphene
\cite{guinea} and thus essentially effect on its electronic
properties. Also, this ``topological protectorate'' of zero-energy
Landau level can explain why it is narrower than the higher-energy
levels as was recently observed experimentally \cite{jos}. The
ripples in bilayer graphene has been already observed
experimentally \cite{meyer2} but it is still not clear whether
this leads to formation of the zero-energy states, similar to the
case of single layer, or not. Here we give a positive answer on
this question based on a very general topological consideration.

\section{Formulation of the problem}

The corrugation leads to important consequences for the electronic
structure of graphene. The nearest-neighbor hopping integral
$\gamma$ turns out to be fluctuating due to its dependence on the
deformation tensor \cite{nelson}
\begin{equation}
\overline{u}_{ij}=\frac{1}{2}\left( \frac{\partial u_{i}}{\partial x_{j}}+%
\frac{\partial u_{j}}{\partial x_{i}}+\frac{\partial u_{k}}{\partial x_{i}}%
\frac{\partial u_{k}}{\partial x_{j}}+\frac{\partial h}{\partial x_{i}}\frac{%
\partial h}{\partial x_{j}}\right)   \label{metric}
\end{equation}
where $h$ is the displacement in the direction perpendicular to
graphene plane, $x_{i}=\left( x,y\right) $ \ are coordinates in
the plane and $u_{i}$ are corresponding components of the
displacement vector:
\begin{equation}
\gamma =\gamma _{0}+\left( \frac{\partial \gamma }{\partial \overline{u}_{ij}%
}\right) _{0}\overline{u}_{ij}.  \label{depend}
\end{equation}
Taking into account this inhomogeneity in a standard tight-binding
description of the electronic structure of graphene \cite{r4} one
can obtain an effective Dirac-like Hamiltonian describing electron
states near the conical $K$-point:
\begin{equation}
H=v_{F}\mathbf{\sigma }\left( -i \hbar \nabla -\mathcal{A}\right)
\end{equation}
where $v_{F}=\sqrt{3}\gamma _{0}a/2\hbar$ and $\mathcal{A}$ is the
``vector potential'' connected with the deviations of the hopping
parameters $\gamma _{i}$ from their unperturbed value $\gamma
_{0}$:
\begin{eqnarray}
\mathcal{A}_{x} &=&\frac{1}{2v_{F}}\left( \gamma _{2}+\gamma
_{3}-2\gamma
_{1}\right),  \nonumber \\
\mathcal{A}_{y} &=&\frac{\sqrt{3}}{2v_{F}}\left( \gamma
_{3}-\gamma _{2}\right) ,
\end{eqnarray}
where the nearest neighbors with vectors $\left(
-a/\sqrt{3},0\right) ;\left( a/2\sqrt{3},-a/2\right) ;\left(
a/2\sqrt{3},a/2\right) $ are labelled 1,2, and 3, correspondingly,
$a$ is the lattice constant \cite{r5}. This means that the
flexural fluctuations act on the electronic structure near the
$K$-point as an Abelian gauge field which is equivalent to the
action of a random magnetic field. Thus, the bending of graphene
violates the time-reversal symmetry for a given valley; of course,
the Umklapp processes between $K$ and $K'$ points will restore
this symmetry. As was suggested in Ref. \onlinecite{morozov} these
effective magnetic fields might be responsible for suppression of
the weak localization effects in graphene.

Whereas a smooth deformation of the graphene sheets produces the
gauge field similar to electromagnetic one, different topological
defects in graphene inducing inter-valley (Umklapp) processes can
be considered as sources of a non-Abelian gauge field;
corresponding analogy with gravitation was discussed in Refs.
\onlinecite{crespi,vozmed}.

The bilayer graphene in a simplest approximation can be considered
as a zero-gap semiconductor with parabolic touching of the
electron and hole bands described by the single-particle
Hamiltonian \cite{natphys,falko1}
\begin{equation}
H=\left(
\begin{array}{cc}
0 & -\left( p_x-ip_y\right) ^2/2m \\
-\left( p_x+ip_y\right) ^2/2m & 0
\end{array}
\right)   \label{bilayer}
\end{equation}
where $p_i=-i\hbar \partial /\partial x_i - \mathcal{A}_{i}$ are electron
momenta operators and $m\simeq 0.054m_e$ is the effective mass,
$m_e$ being the free-electron mass. This description is accurate
at the energy scale larger than few meV, otherwise a more
complicated picture including trigonal warping takes place
\cite{falko1}; we will restrict ourselves only by the case of not
too small doping when the approximate Hamiltonian (\ref{bilayer})
works. Two components of the wave function are originated from
crystallographic structure of graphite sheets with two carbon
atoms in the sheet per elementary cell. There are two touching
points per Brillouin zone, $K$ and $K^{\prime }$. For smooth
enough external potential, no Umklapp processes between these
points are allowed and thus they can be considered independently.

We will proof that the zero-energy states in the case of bilayer
found by exact solution for the case of homogeneous magnetic field
\cite{falko1} are topologically protected and their number is
determined only by the total flux per sample, irrespective to
whether the field is homogeneous or not, exactly as in the case of
the single-layer \cite{r2,r5,kostya}.

\section{Results and discussion}

The proof is based on the theory of elliptic operators and on the
Atiyah-Singer index theorem. Let us remind first some facts about
it.

Let $X$ be a smooth compact manifold, $E$ and $E'$ smooth complex
bundles over $X$
(we shall use everywhere the word ``smooth'' in the sense of
``infinitely differentiable'').
Let $D$ be a smooth linear differential operator
of order $m$ acting from $C^{\infty}(E)$ to $C^{\infty}(E')$ where
$C^{\infty}(E)$ is the vector space of smooth sections of $E$.
Here smoothness of operator is regarded as smoothness of its
coefficients in any smooth local coordinates.

In local coordinates $(x^i)$ on $X$ the highest-order terms of $D$
have a form $\sum a^{i_1 \ldots i_m}(x) \frac{\partial}{\partial
x^{i_1}} \ldots \frac{\partial}{\partial x^{i_m}}$. Let us
consider the expression $\sum a^{i_1 \ldots i_m}(x) \xi_{i_1}
\ldots \xi_{i_m}$, $\xi \in T^{*}X$, $T^{*}X$ being the cotangent
bundle of $X$. It is independent on the choice of local
coordinates and defines the homomorphism of vector bundles
$\pi_{*}E \to \pi_{*}E'$, which is homogeneous of degree $m$ by
$\xi$. Here $\pi_{*}E$, $\pi_{*}E'$ are the liftings of the
bundles $E$, $E'$ to $T^{*}X$ (see the commutative diagrams below;
vertices on the diagrams are smooth manifolds and arrows are
smooth maps).
\[
\diagram
\pi_{*}E \dto \rto^{\pi_{*}} & E \dto
\\
T^{*}X \rto^{\pi} & X
\enddiagram
\quad \quad
\diagram
\pi_{*}E \drto \rrto^{\sigma(D)} & & \pi_{*}E' \dlto
\\
& T^{*}X &
\enddiagram
\]
This homomorphism $\pi_{*}E \to \pi_{*}E'$ is called
\textit{symbol} $\sigma(D)$ of differential operator $D$. The
latter is called \textit{elliptic} if $\sigma(D)$ is invertible
outside zero section of $T^{*}X$ (that is invertible at $\xi \neq
0$ in local coordinates).

Elliptic operators have a good behavior \cite{AS1}: if $D \colon
C^{\infty}(E) \to C^{\infty}(E')$ is a smooth elliptic operator
then
\begin{itemize}
    \item All distributional solutions of $D$ are smooth.
    \item
$\Ker D$ (the space of solutions of the equation $D\psi =0$) and
$\Coker D$ (the factor-space of $C^{\infty}(E')$ by the image $\{D\psi\}$ of
$D$) are finite dimensional.
    \item $\ind D \triangleq \dim \Ker D - \dim\Coker D$ depends only on the symbol of $D$;
    moreover, $\ind D$ depends only on the homotopy class of symbol in the space of continuous invertible symbols of a given order.
\end{itemize}

The rough idea of the proof is the following. We consider the
two-periodic case (a justification of this choice will be
discussed below). So a wave function is the section of complex
linear bundle $E$ over two-dimensional torus $X$. The vector
potential $\mathcal{A}$ defines the connection $\nabla_j =
\frac{\partial}{\partial x_j}-\frac{i}{\hbar}\mathcal{A}_j$ on
$E$; its flux $\hbar^{-1}\int_{X}\d \mathcal{A}$ defines the
bundle $E$ up to isomorphism.

The differential operators $p_x \pm i p_y$ are elliptic operators
acting on $C^{\infty}(E)$. They are conjugated so the co-kernel of
$p_x + i p_y$ is isomorphic to the kernel of $p_x - i p_y$ and
vice versa. Therefore the difference of dimensions of the kernels
of $p_x + i p_y$ and $p_x - i p_y$
is equal to the index of operator $p_x + ip_y $.
The same is valid for the squares of these operators
thus $\dim \Ker (p_x + i p_y)^2 - \dim \Ker (p_x - i p_y)^2 = \ind (p_x + i p_y)^2$.

The desired result follows just from the fact that the index of
the composition of elliptic operators is equal to the sum of their
indices \cite{Palais}, so we have $\ind (p_x + i p_y)^2 = 2\ind
(p_x + i p_y)$.

According to the Atiyah-Singer theorem the index of operator $p_x
+ i p_y$ depends on only the symbol of $p_x + i
p_y$ determined by the integer number $N=(2\pi
\hbar)^{-1}\int_{X}\d \mathcal{A}$
and does not depend on the choice of the field $\mathcal{A}$ for a
given number of this integral. We obtain $\ind (p_x + i p_y)=N$
from purely topological considerations, replacing operator $p_x +
i p_y$ by other operator with the same symbol and known index. In
physical terms, $N$ is the total flux of the (pseudo)magnetic
field per torus in the units of the flux quantum.

It is worth to stress that the ``vector potentials'' $\mathcal{A}$
are assumed to be in our proof not very smooth but just
continuously differentiable which makes the result rather general.

The choice of the torus can be justified by standard arguments
used at the introduction of the Born - von Karman periodic
boundary conditions in solid state theory \cite{maradudin}.
Namely, the total number of zero-energy modes, assuming that $N
\neq 0$ is proportional to the total number of atoms in the
sample, $N_0$. At the same time, if one replaces ``realistic''
boundary conditions by the periodic ones the total density of
states can be changed by a quantity proportional to the number of
edge atoms, that is, $\sqrt{N_0}$. This means that the total
number of states with the energy {\it close} to zero should be, in
the limit of large crystallite, correctly described by the
periodic case, that is, the case of torus. Note that the torus has
zero Gaussian curvature which physically means absence of
topological defects, such as disclinations (pentagons of heptagons
in the original hexagonal lattice) \cite{crespi,vozmed}.

The physical consequences are straightforward: as well as for the
case of the single-layer graphene, for the case of bilayer (i)
corrugations can result in the appearance of the mid-gap states
\cite{guinea} and (ii) pseudomagnetic fields due to corrugations
will not broaden the zero-energy Landau level in the case of
quantum Hall effect \cite{jos}. It would be very interesting to
check experimentally the second statement by measurements of the
quantum Hall activation gaps for the bilayer graphene, similar to
Ref. \onlinecite{jos} for the case of single layer.

A mathematical proof of the statement is presented in the
Appendix.

\section*{Acknowledgements}
We are grateful to Maria Vozmediano for helpful discussions.
M.I.K. acknowledges financial support from Stichting voor
Fundamenteel Onderzoek der Materie (FOM), the Netherlands. The
work of M.F.P. was partially supported by a RFBR grant
08-01-00029, Russia.

\section*{Appendix}


Let $\Gamma$ be a lattice in the two-dimensional Euclidean space
$\R^2$, $X=\R^2 \mod \Gamma$ be the two-dimensional torus, $E$ be
a smooth linear complex vector bundle over $X$ with structure
group $U(1)$. Let we have $C^1$-connection on $E$, that is
corresponding covariant derivatives are written as $\nabla_x =
\frac{\partial}{\partial x}-i A_x$, $\nabla_y =
\frac{\partial}{\partial y}-i A_y$, where $A_x$, $A_y$ are
\textit{continuously differentiable} real functions in local
coordinates $(x,y)$ on $X$ (these coordinates we choose as usual
coordinates on universal covering $\R^2$ of $X$, so they are
defined up to addition of vectors from $\Gamma$).

Although the connection form $A=A_x\d x+A_y\d y$ depends on the
choice of local coordinates, its curvature $\d A$ is globally
defined on $X$. We can integrate $\d A$ over $X$; this integral
depends only on isomorphism class of $E$ and does not depend on the choice of
connection on $E$. Let
\begin{equation}
    N(E)=(2\pi)^{-1}\int_{X}\d A;
\end{equation}
this number must be integer.

Following Ref. \onlinecite{AS1}, we consider Hilbert spaces
$H_s(E)$ of those distributional sections $u$ of $E$ for which $Du
\in L_2(X)$ for all differential operators $D \colon C^{\infty}(E)
\to C^{\infty}(1_X)$ with smooth coefficients, and of order $\leq
s$. Here $C^{\infty}(E)$ is the space of smooth sections of $E$,
$1_X$ is the trivial linear complex bundle over $X$. The Hermitian
product in $H_s(E)$ can be defined at $s=0$ as $\left\langle
u,v\right\rangle_0 = \int_X \left\langle u,v\right\rangle\d x \d
y$, at $s>0$ -- as $\left\langle u,v\right\rangle_s = \int_X
\left\langle \Delta^s u,v\right\rangle\d x \d y$. Here
$\Delta=1+D^{*}D$, $D\colon C^{\infty}(E) \to C^{\infty}(E \otimes
T^{*}X)$ is the covariant derivative given by some fixed smooth
connection on $E$ (precise choose of this connection is irrelevant
for our aims).

We can consider differential operators $ P^{\pm}=\nabla_x \pm
i\nabla_y$ as continuous linear operators from $H_s(E)$ to
$H_{s-1}(E)$ at $s\leq 2$; let us denote these linear operators as
$ P_s^{\pm}$. Similarly, we can consider differential
operators $Q^{\pm} = \left( \nabla_x \pm i \nabla_y \right)^2$ as
continuous linear operators $Q_2^{\pm}$ from $H_2(E)$ to $H_0(E)$.

\textbf{Theorem.}
\begin{equation}\label{eq:teo}
    \begin{aligned}
        \dim \Ker  P_1^{+} - \dim \Ker  P_1^{-} = N(E), \\
        \dim \Ker Q_2^{+} - \dim \Ker Q_2^{-} = 2N(E).
    \end{aligned}
\end{equation}

\textbf{Remark.} If the connection (that is the functions $A_x$, $A_y$)
is \textit{smooth} then all distributional solutions of the operators
$ P_1^{\pm}$, $Q_2^{\pm}$ are also smooth \cite{AS1}, and for the
differential operators $ P^{\pm}$, $Q^{\pm}$ acting on
$C^{\infty}(E)$ we have from Eq.(\ref{eq:teo})
\begin{gather*}
            \dim \Ker  P^{+} - \dim \Ker  P^{-} = N(E), \\
            \dim \Ker Q^{+} - \dim \Ker Q^{-} = 2N(E).
\end{gather*}

\textbf{Proof.} Note that in our case when $X$ is the
two-dimensional torus the cotangent bundle $\pi \colon T^{*}X \to
X$ is trivial two-dimensional real bundle over $X$ and can be
identified with the trivial linear complex bundle  $X \times \CC
\to X$. So lift of the bundle $E$ over $X$ to the bundle
$\pi_{*}E$ over $T^{*}X$ can be identified with linear complex
bundle $E \times \CC \to X \times \CC$. At this identification
symbols $\sigma^{\pm} \colon \pi_{*}E \to \pi_{*}E$ of the
operators $ P^{\pm}$ become the following form: for $e \in E$,
$\xi \in \CC$ we have $\sigma^{+}(e,\xi)=(\xi e, \xi)$,
$\sigma^{-}(e,\xi)=(\bar{\xi} e, \xi)$, that is the fiber over a
point of $T^{*}X$ is multiplied by the complex number
corresponding to this cotangent vector in the case $\sigma^{+}$,
and on the conjugate to this complex number in the case
$\sigma^{-}$.

The composition $\sigma^{+} \sigma^{-}\colon \pi_{*}E \to
\pi_{*}E$, $\sigma^{+} \sigma^{-}(e,\xi)=(|\xi|^2 e, \xi)$
coincides with the identity $\id \colon \pi_{*}E \to \pi_{*}E$ on
the unit sphere bundle of $T^{*}X$. So $[\sigma^{+}] +
[\sigma^{-}]=0$, where $[\sigma]$ is the class of $\sigma$ in the
group $K\left(T^{*}X\right)$ where $K$ denotes $K$-theory with
compact supports (a description of this variant of $K$-theory is
contained in Ref. \onlinecite{AS1}).

Applying ``topological index'', that is, homomorphism $\ind \colon
K\left(T^{*}X\right) \to \Z$ constructed by Atiyah and
Singer\cite{AS1}, to this equality, we get
\[
\ind [\sigma^{+}] + \ind [\sigma^{-}]=0.
\]

The operators $ P_s^{\pm} \colon H_s(E) \to H_{s-1}(E)$ are
Fredholm since symbols $\sigma^{\pm}$ are invertible outside the
zero section of $T^{*}X$ (Ref. \onlinecite{AS1}). $\ind  P_s^{\pm}
\triangleq \dim \Ker  P_s^{\pm} - \dim \Coker  P_s^{\pm}$ depends
only on $[\sigma^{\pm}]$ and are independent of the choice of
$s\leq 2$ and connection field $A$ (but of course they depend on
$N(E)$, which define the isomorphism class of $E$): $\ind
 P_s^{\pm} = \ind [\sigma^{\pm}]$ (Ref. \onlinecite{AS1}).

Note that at $s\geq 1$ for $u, v \in H_s(E)$ we have $u\bar{v} \in
H_s(1_X)$. So $\int_X(u\bar{v})_x \d x \d y = \int_X(u\bar{v})_y
\d x \d y =0$, and $\left\langle P_1^{+}u, v\right\rangle_0 +
\left\langle u, P_1^{-}v\right\rangle_0 =0$ for any $u, v \in
H_1(E)$. Identifying $\Coker P_1^{\pm}$ with the orthogonal
complement of $\Image P_1^{\pm}$ in $H_0(E)$, we obtain
\[
\left\{
\begin{aligned}
    \Ker  P_1^{-} = \left( \Coker  P_1^{+} \right) \cap H_1(E) \\
    \Ker  P_1^{+} = \left( \Coker  P_1^{-} \right) \cap H_1(E)
\end{aligned}
\right.
\]
Hence
\[
\left\{
\begin{aligned}
    \ind  P_1^{+} \leq \dim \Ker  P_1^{+} - \dim \Ker  P_1^{-}  \\
    \ind  P_1^{-} \leq \dim \Ker  P_1^{-} - \dim \Ker  P_1^{+}
\end{aligned}
\right.
\]
where every of these inequalities become equality if the
co-kernel of the corresponding operator contains in $H_1(E)$.
However, $\ind  P_1^{+} + \ind  P_1^{-} = \ind
[\sigma^{+}] + \ind [\sigma^{-}]=0$, so both inequalities should
be equalities, and we obtain $\Coker  P_1^{\pm} \subset
H_1(E)$, and $\Ker  P_1^{\pm} \cong \Coker  P_1^{\mp}$.
Hence,
\begin{equation}
    \dim \Ker  P_1^{+} - \dim \Ker  P_1^{-} = \ind  P_1^{+} = \ind [\sigma^{+}].
\label{eq:ind1}
\end{equation}

Repeating this consideration almost literally for the operators
$Q_2^{\pm} = \left( \nabla_x \pm i \nabla_y \right)^2 =
 P_1^{\pm} P_2^{\pm}$ acting from $H_2(E)$ to $H_0(E)$,
and using the fact that index of the composition of Fredholm
operators is equal to the sum of their indices \cite{Palais}, we
have
\begin{gather}
        \Coker Q_2^{\pm} \subset H_2(E), \notag \\
        \Ker Q_2^{\pm} \cong \Coker Q_2^{\mp}, \notag \\
        \dim \Ker Q_2^{+} - \dim \Ker Q_2^{-} = \ind Q_2^{+} = \ind ( P_1^{+} P_2^{+}) = \notag \\
         = \ind  P_1^{+} + \ind  P_2^{+} = 2\ind [\sigma^{+}].
\label{eq:ind2}
\end{gather}

We present below an explicit calculation of the value of $\ind
[\sigma^{+}]$ based on one famous theorem from algebraic geometry.
But for more clarity we start with a simple reasoning showing the
proportionality of $\ind [\sigma^{+}]$ to $N(E)$.

Let us see on the construction of complex bundles over $X$. Let
$F$ be a $U(n)$-vector bundle over $X$. Cut out the disk $B^2$
from the torus $X$. Since the disk is contractible, the
restriction of $F$ to $B^2$ is trivial. $X-B^2$ is homotopically
equivalent to the wedge product of two circles, and $U(n)$ is
connected, so the restriction of $F$ to $X-B^2$ is trivial, too.
Thus, the isomorphism class of the bundle $F$ is uniquely defined
by its dimension $n$ and by the homotopical class of the map
$\varphi \colon S^1 = \partial B^2 \to U(n)$ gluing together two
these trivial bundles. This homotopical class is defined by the
degree of the map $\det \cdot \varphi \colon S^1 \to U(1)$, where
$\det \colon U(n) \to U(1)$ is the determinant homomorphism. The
sum of degrees corresponds to the Whitney sum of a bundles over
$X$, so $K(X)=\Z \oplus \Z$. Here $K(X)$ is Abelian group
generated by elements $[F]$ with relations $[F\oplus F'] =
[F]+[F']$ for all complex bundles $F$, $F'$ over $X$; detailed
description of $K$-theory is contained in Ref.\onlinecite{Atiyah}.

Particularly, the isomorphism class of a linear bundle $E$ is defined by the integer
$N(E) = \deg (\det\cdot \varphi) = (2\pi)^{-1}\int_{X}\d A$,
and for the class $[E]$ of $E$ in $K(X)$ we have
\begin{equation}\label{eq:N}
[E]-1=N(E)([E_1]-1),
\end{equation}
where $E_1$ is a linear bundle over $X$ for which $N(E_1)=1$.

If $E$ is trivial then choosing trivial connection we obtain that
$\Ker P^{+}$ is the space of holomorphic functions on torus and
$\Ker P^{-}$ is the space of anti-holomorphic functions on torus.
Both these spaces contain only constants and are 1-dimensional, so
$\ind P^{+}=0$ in this case. Taking into account that $\sigma^{+}$
is the image of $[E] \in K(X)$ at the Thom isomorphism $K(X) \to
K(X \times \CC)$ (the description of this isomorphism see in Ref.
\onlinecite{AS1}), from (\ref{eq:N}) we obtain that $\ind
[\sigma^{+}]$ is proportional to $N(E)$.

To calculate the coefficient of this proportionality, moreover, to
calculate the value of $\ind [\sigma^{+}]$, let us replace $E$ by
the other bundle of the same class in $K(X)$, and replace
$ P_s^{+}$ by the other operator of the same symbol class in
$K(T^{*}X)$ (the index of the operator does not change at such a
replacement).

Consider the torus $X$ as algebraic curve, with local complex
coordinate $z=x+iy$. Choose holomorphic line bundle $F$ over $X$
isomorphic to $E$ in smooth category, that is such that
$N(F)=N(E)$ (for example, we can take divisor on $X$ consisting of
a point $z_0 \in X$ of the multiplicity $N(E)$, and turn from the
divisor to corresponding holomorphic line bundle by the way
described in Ref. \onlinecite{KS}).

Consider now the differential operator $\bar{\partial} \colon
C^{\infty}(\F^{0,0}) \to C^{\infty}(\F^{0,1})$, $\bar{\partial}=
\frac{\partial}{\partial \bar{z}}= \frac{\partial}{\partial x} + i
\frac{\partial}{\partial y}$, where $\F^{0,k}$ is the bundle of
differential forms on $X$ of type $(0,k)$ with coefficients in
$F$. Since the complex cotangent bundle of $X$ is trivial and
linear, we have $\F^{0,0} \cong \F^{0,1} \cong F$, so the symbol
of the operator $\bar{\partial}$ is coincide with $\sigma^{+}$.
Hence $\ind [\sigma^{+}] = \ind \bar{\partial}$. However, $\ind
\bar{\partial}$ is equal to the Euler characteristic $\chi(X,F)$
of the sheaf of germs of holomorphic sections of $F$. We can
compute $\chi(X,F)$ using the Riemann-Roch-Hirzebruch theorem
\cite{Hirz}. For a curve $X$ and linear bundle $F$ this theorem
yield $\chi(X,F) = c_1(F)[X] +1-g$, where $g$ is the genius of the
curve $X$ and $c_1(F) \in H^2(X;\Z)$ is the first Chern class of
the bundle $F$. In our case $g=1$, $c_1(F)[X]=c_1(E)[X]=
(2\pi)^{-1}\int_{X}\d A = N(E)$, so we obtain the final formula
\begin{equation}\label{eq:ind}
    \ind [\sigma^{+}]=N(E).
\end{equation}

\noindent Substituting (\ref{eq:ind}) to
(\ref{eq:ind1})-(\ref{eq:ind2}), we obtain the assertion of the
Theorem.

\end{document}